\documentclass[twocolumn]{jpsj3}

\usepackage{color}
\usepackage{times}
\usepackage{graphicx}

\title{Effect of Spin-Orbit Coupling on Kondo Phenomena
in $f^7$-Electron Systems}

\author{Takashi Hotta}

\inst{Department of Physics, Tokyo Metropolitan University,
Hachioji, Tokyo 192-0397, Japan}

\recdate{\today}

\abst{
In order to promote our basic understanding on the Kondo behavior
recently observed in europium compounds,
we analyze an impurity Anderson model with seven $f$ electrons
at an impurity site by employing a numerical renormalization group method.
The local part of the model consists of Coulomb interactions among $f$ electrons,
spin-orbit coupling $\lambda$, and crystalline electric field (CEF) potentials,
while we consider the hybridization $V$ between local $f$ electrons and
single-band conduction electrons with $a_{\rm u}$ symmetry.
For $\lambda=0$, we observe the underscreening Kondo behavior
for appropriate values of $V$, characterized by the entropy change
from $\ln 8$ to $\ln 7$,
in which one of seven $f$ electrons is screened by conduction electrons.
When $\lambda$ is increased, we obtain two types of behavior
depending on the values of $V$.
For large $V$, we find the entropy release of $\ln 7$ at low temperatures,
determined by the level splitting energy due to the hybridization.
For small $V$, we also observe the entropy change from $\ln 8$ to $\ln 2$
by the level splitting due to the hybridization, but at low temperatures,
$\ln 2$ entropy is found to be released, leading to the Kondo effect.
We emphasize that the Kondo behavior for small $V$ is observed
for realistic values of $\lambda$ in the order of $0.1$ eV.
We also discuss the effect of CEF potentials and the multipole properties
in the Kondo behavior found in this paper.
}

\begin{document}
\maketitle

\section{Introduction}

Research on heavy-electron systems has been one of central topics
in the field of strongly correlated electron physics.\cite{Kondo40,ICHE2010,SCES2013}
The origin of such heavy-electron state has been understood
on the basis of quantum critical phenomena,
\cite{review1,review2,review3,review4,review5,review6,review7}
emerging in the competing region of itinerant properties of $f$ electrons
due to the Kondo effect~\cite{Kondo1,Kondo2a,Kondo2b}
and localized nature due to the Ruderman-Kittel-Kasuya-Yosida (RKKY)
interaction.\cite{RKKY1,RKKY2,RKKY3}

The Kondo effect has been first discovered as resistance minimum
phenomena in good metals such as Cu, Ag, and Au with a small amount of
magnetic impurities such as Mn and Fe.
In the dilute system, it has been widely recognized that Kondo effect
occurs when the singlet state is formed from local magnetic moment
due to the antiferromagnetic (AF) coupling with conduction electrons.\cite{Yosida}
As for the original problem of resistance minimum phenomena
in metals with magnetic impurities,
Kondo has actually demonstrated it by quantum-mechanical calculations
for scattering amplitude of electrons due to magnetic impurities.
\cite{Kondo1,Kondo2a,Kondo2b}

On the other hand, there appears long-range interaction between localized
electrons via conduction electrons.
This is called the RKKY interaction,\cite{RKKY1,RKKY2,RKKY3}
which enhances the localized nature of $f$ electrons.
As a result of the competition between the Kondo effect and the RKKY
interactions, quantum criticality appears between AF and metallic phases,
when we control the coupling constant $\rho J_{\rm cf}$,
where $J_{\rm cf}$ denotes the AF coupling between localized and
conduction electrons and $\rho$ is the density of states at the Fermi level.
The picture has been summarized in the Doniach's phase diagram,\cite{Doniach}
which has been a guiding principle in the heavy-electron physics to
discover the unconventional superconductivity mediated by
quantum critical fluctuations.

Along this line, the heavy-electron phenomena and unconventional superconductivity
have been intensively and extensively investigated in the Ce compounds
with one $f$ electron for Ce$^{3+}$ ion,
since the pioneering discovery of
superconductivity in CeCu$_2$Si$_2$.\cite{Steglich}
Recently, the research on heavy-electron state and superconductivity in
Yb compounds has been activated on the basis 
of the electron-hole conversion picture of Ce$^{3+}$.
\cite{Nakatsuji1,Nakatsuji2,Nakatsuji3}
It has been claimed that this material exists just on the
quantum critical point at ambient pressure.
For the case of two $f$ electrons, there is a long history in the research
of U compounds.
In fact, after the discovery of heavy-electron superconductivity
in CeCu$_2$Si$_2$,
superconductivity has been found in U compounds
such as UBe$_{13}$,\cite{UBe13}
UPt$_3$,\cite{UPt3} URu$_2$Si$_2$,\cite{URu2Si2}
UPd$_2$Al$_3$,\cite{UPd2Al3} and UNi$_2$Al$_3$.\cite{UNi2Al3}
In a recent decade, superconductivity and magnetism in
Pr compounds such as PrOs$_2$Sb$_{12}$,\cite{Bauer}
PrPb$_3$,\cite{Onimaru1,Onimaru2} and
PrT$_2$X$_{20}$ (T: transition metal, X=Al and Zn)
\cite{Onimaru3,Onimaru4,Sakai1,Sakai2,Matsubayashi,Tsujimoto}
have been actively investigated.
On the basis  of the electron-hole conversion picture of Pr$^{3+}$,
it is also interesting to note Tm$^{3+}$ with twelve $f$ electrons.
In Tm$_5$Rh$_6$Sn$_{18}$, superconductivity has been
found with $T_{\rm c}=2.2$K.\cite{Kase}
Peculiar reentrant properties have been considered to be
related to the coexistence of magnetism and superconductivity.

When we look back over the history of heavy-electron materials,
we immediately notice that elements around both ends of lanthanide series
have been focused thus far.
However, recently, the heavy-electron state in Eu compounds,
just at the center in lanthanide series, have attracted renewed attention.
\cite{Mitsuda1,Hiranaka,Mitsuda2,Nakamura1,Nakamura2}
In the divalent ion of europium, seven electrons are accommodated
in the $4f$ orbitals, while in the trivalent ion of europium, we find
six $4f$ electrons.
In an $LS$ coupling scheme, due to the Hund's rules,
we obtain $J=S=7/2$ and $L=0$ for Eu$^{2+}$, where
$J$, $S$, and $L$ indicate, respectively, total angular momentum,
total spin momentum, and total orbital momentum.
On the other hand, we find $J=0$ with $S=L=3$ for Eu$^{3+}$,
which is non-magnetic.
The difference in magnetic properties between Eu$^{2+}$ and Eu$^{3+}$
ions are significant, but in general, the valence fluctuations easily occur
in Eu compounds, since the energy difference between two valence states
has been known to be small.
Thus, the valence fluctuation is one of key issues in Eu compounds.
\cite{Mitsuda2}

Quite recently, in Eu$_2$Ni$_3$Ge$_5$ and EuRhSi$_3$,
heavy-electron states have been claimed to be observed in the
temperature dependence in the measurement of resistivity.\cite{Nakamura2}
At ambient pressure, those compounds are in the AF state
at low temperatures, but under pressure such as several GPa,
the AF state is suppressed and three characteristic temperatures have
been observed in the resistivity.
The highest one was assigned as $T_{\rm v}$, which is the valence transition
temperature.
Other two low temperatures were considered as Kondo temperatures.
These results were claimed to be quite similar to those of Ce compounds.

At a first glance, it seems to be difficult to accept the similarity between
Ce and Eu compounds, but it is easy to hit upon an idea on the basis
of a $j$-$j$ coupling scheme, where $j$ denotes the total angular momentum
of one $f$ electron.
Namely, in the $j$-$j$ coupling scheme, among seven $f$ electrons,
six electrons fully occupy the sextet of $j=5/2$,
whereas one electron is accommodated in the octet of $j=7/2$,
as shown in Fig.~1(a).
Since six electrons in the sextet do not contribute to electronic properties,
one $f$ electron plays a main role for the Kondo effect,
leading to the similar behavior as that of Ce compound.
This idea has been also emphasized by the present author for
the explanation of active quadrupole degrees of freedom in Gd compound
with seven $f$ electrons in the trivalent ion state of gadolinium.\cite{Hotta1}

Here we cast a naive question:
Are there any problems to use the $j$-$j$ coupling scheme
for the Eu compounds?
In fact, in a textbook of solid state physics,
it is standard to use the $LS$ coupling scheme.
Namely, first we construct the many $f$ electron state
characterized by $S$ and $L$ due to the so-called Hund's rules.
Then, we consider the effect of spin-orbit coupling
by including the term of $\Lambda {\mib L}\cdot{\mib S}$
with a spin-orbit coupling $\Lambda$,
leading to the multiplet characterized by $J$.
For seven $f$ electrons, as mentioned above, first we obtain the state
with $S=7/2$ and $L=0$ from the Hund's rules, as shown in Fig.~1(b).
Since $L$ is zero, the ground state is characterized by $J=S=7/2$.

\begin{figure}[t]
\centering
\includegraphics[width=8.0truecm]{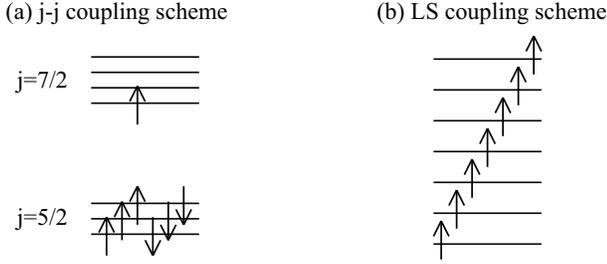}
\caption{
Electron configurations of the seven $f$-electron state
for (a)  the $j$-$j$ coupling scheme
and (b) the $LS$ coupling scheme.
Note that up and down arrows denote, respectively,
pseudo-spin up and down electrons in (a),
while in (b), they indicate real-spin up and down electrons.
The pseudo-spin state is defined through the time-reversal relation.
}
\end{figure}

For rare-earth compounds, in general, the magnitude of Hund's rule interaction
among $f$ orbitals is a few eV, while the spin-orbit coupling
takes a value of a few thousand Kelvins.
If we should take one of the two limiting situations, it is better to choose
the $LS$ coupling scheme, as readers have learned in the standard textbook.
In this sense, it seems to be difficult to understand the Kondo effect in
the Eu compound on the basis of the $j$-$j$ coupling scheme.

However, we should remark that both of the spin-orbit coupling and
the Hund's rule interaction are finite in actual materials and
the actual situation is always in the middle of the $LS$ and $j$-$j$ coupling schemes.
Namely, the wave function of the many $f$ electron state is the mixture of
those in the $LS$ and $j$-$j$ coupling schemes.
In order to discuss the Kondo effect in the Eu compound even qualitatively,
it is essential to consider both of the spin-orbit coupling and
the Hund's rule interaction.
This point has been emphasized in the research of quadrupole susceptibility
in Gd compounds.\cite{Hotta1}

In this paper, we analyze the seven-orbital Anderson model by employing
a numerical renormalization group technique.
The local term contains Coulomb interaction, spin-orbit coupling,
and crystalline electric field (CEF) potential terms.
Here we introduce the Hund's rule interaction $U$,
the spin-orbit coupling $\lambda$, and the CEF potential $W$.
As for conduction electrons, we include only one conduction band
with a$_{\rm u}$ symmetry.
Readers may consider that a trivial result of the underscreening Kondo
effect is obtained, but such a well-known result is observed only in the
$LS$ coupling limit.
When we increase the spin-orbit coupling in this situation,
we confirm the Kondo bahavior similar to that in the Ce compound,
as expected from the $j$-$j$ coupling scheme.
An important finding in this paper is that the Kondo effect
similar to the case of $n=1$ occurs for a realistic value
of spin-orbit coupling even in the case of $n=7$,
where $n$ denotes local $f$-electron number.
When we explicitly include the cubic CEF potentials,
we discuss the multipole properties and the Kondo behavior
similar to Ce compounds.

The organization of this paper is as follows.
In Sec.~2, we provide the local $f$-electron Hamiltonian
and briefly review the change of the seven $f$-electron
states between the $LS$ and $j$-$j$ coupling schemes
by evaluating the Curie constant.
We emphasize that the transition region between them
just corresponds to the situation of actual $f$-electron materials.
In Sec.~3, we introduce the impurity Anderson model
to discuss the Kondo phenomenon for the case of $n=7$.
We also briefly explain the method used in this paper and
provide the definition of multipole susceptibilities.
In Sec.~4, we exhibit our numerical results and discuss
how an entropy changes with the decrease of a temperature.
First we clearly show the underscreening Kondo effect
in the $LS$ coupling scheme for $\lambda=W=0$.
Then, we investigate the effect of the spin-orbit coupling on
the underscreening Kondo behavior for $W=0$.
For small $V$ and $\lambda/U$ in the order of $0.1$,
we confirm the Kondo behavior characterized by
the entropy release of $\ln 2$, as found in the Ce compound.
Then, we also investigate the Kondo behavior for $W \ne 0$
and discuss the multipole susceptibility to confirm
that the relevant multipole is dipole in the present Kondo effect.
Finally, in Sec.~5, we provide a few comments on future problems
and summarize this paper.
Throughout this paper, we use such units as $\hbar=k_{\rm B}=1$.

\section{Local $f$ Electron State}

\subsection{Local Hamiltonian}

First, we define the local $f$-electron Hamiltonian as
\begin{equation}
\label{Hloc}
\begin{split}
  H_{\rm loc} &=\sum_{m_1 \sim m_4}\sum_{\sigma,\sigma'}
  I_{m_1m_2,m_3m_4}
  f_{m_1\sigma}^{\dag}f_{m_2\sigma'}^{\dag}
  f_{m_3\sigma'}f_{m_4\sigma}\\
 &+ \lambda \sum_{m,\sigma,m',\sigma'}
   \zeta_{m,\sigma;m',\sigma'} f_{m\sigma}^{\dag}f_{m'\sigma'}\\
 &+ \sum_{m,m',\sigma} B_{m,m'}
      f_{m \sigma}^{\dag} f_{m' \sigma},
\end{split}
\end{equation}
where $f_{m\sigma}$ denotes the annihilation operator
for local $f$ electron with the spin $\sigma$ and $z$-component $m$
of the angular momentum $\ell=3$,
$\sigma=+1$ ($-1$) for up (down) spin,
$I_{m_1m_2,m_3m_4}$ indicates the Coulomb interaction,
$\lambda$ is the spin-orbit coupling,
and $B_{m,m'}$ denotes the CEF potential.

The Coulomb interaction $I$ is known to be expressed as
\begin{equation}
   I_{m_1m_2,m_3m_4} = \sum_{k=0}^{6} F^k c_k(m_1,m_4)c_k(m_2,m_3),
\end{equation}
where $F^k$ indicates the Slater-Condon parameter and
$c_k$ is the Gaunt coefficient.\cite{Slater}
Note that the sum is limited by the Wigner-Eckart theorem to
$k=0$, $2$, $4$, and $6$.
Although the Slater-Condon parameters should be determined
for the material from the experimental results,
we assume the ratio among the Slater-Condon parameters as
\begin{equation}
  F^0=10U,~F^2=5U,~F^4=3U,~F^6=U,
\end{equation}
where $U$ is the Hund's rule interaction among $f$ orbitals.

Each element of $\zeta$ for the spin-orbit coupling is given by
\begin{equation}
\begin{split}
    \zeta_{m,\sigma;m,\sigma}&=m\sigma/2,\\
    \zeta_{m+\sigma,-\sigma;m,\sigma}&=\sqrt{\ell(\ell+1)-m(m+\sigma)}/2,
\end{split}
\end{equation}
and zero for other cases.
The CEF potentials for $f$ electrons from the ligand ions is
given in the table of Hutchings
for the angular momentum $\ell=3$.\cite{Hutchings}
For cubic structure with $O_{\rm h}$ symmetry,
$B_{m,m'}$ is expressed using a couple of CEF parameters,
$B_4^0$ and $B_6^0$, as
\begin{equation}
\begin{split}
    B_{3,3}&=B_{-3,-3}=180B_4^0+180B_6^0, \\
    B_{2,2}&=B_{-2,-2}=-420B_4^0-1080B_6^0, \\
    B_{1,1}&=B_{-1,-1}=60B_4^0+2700B_6^0, \\
    B_{0,0}&=360B_4^0-3600B_6^0, \\
    B_{3,-1}&=B_{-3,1}=60\sqrt{15}(B_4^0-21B_6^0),\\
    B_{2,-2}&=300B_4^0+7560B_6^0,
\end{split}
\end{equation}
Note the relation $B_{m,m'}=B_{m',m}$.
Following the traditional notation,\cite{LLW}
we define $B_4^0$ and $B_6^0$ as
\begin{equation}
    B_4^0=Wx/F(4),~B_6^0=W(1-|x|)/F(6),
\end{equation}
where $x$ and $y$ specify the CEF scheme for the $O_{\rm h}$ point group,
while $W$ determines the energy scale for the CEF potential.
We choose $F(4)=15$ and $F(6)=180$ for $\ell=3$.\cite{Hutchings}

Before proceeding to the results, we provide a comment on the energy scale
for $U$, $\lambda$, and $W$,
Among them, the largest one is $U$ and its magnitude is in the order of 1 eV,
since it denotes the Hund's rule interaction among $f$ orbitals.
The next one is $\lambda$ and its magnitude is in the order of 0.1 eV,
although the precise values depend on the kind of lanthanide and actinide
ions between 0.1 and 0.3 eV.
The smallest one is $W$, since the magnitude is typically in the order of
meV, although the values depend on materials.
In any case, it is reasonable to consider
$U > \lambda \gg W$ and $\lambda/U \sim 0.1$
for actual $f$-electron compounds.

Concerning the energy unit, we define it as a half of
the conduction bandwidth in this paper, as will be mentioned later.
Throughout this paper, we set it as 1 eV, but this value is
realistic as a half of the conduction bandwidth.

\subsection{$LS$ and $j$-$j$ coupling schemes}

In the standard textbook for solid state physics,
it is recommended to employ an $LS$ coupling scheme for
$f$-electron states in the limit of $U \gg \lambda$, while
we should use a $j$-$j$ coupling scheme in the limit of $U \ll \lambda$.
Since the actual situation is found for $\lambda/U \sim 0.1$,
it seems to be better to choose always the $LS$ coupling scheme.
However, as has been emphasized in our previous papers,\cite{Hotta1,Hotta3,Hotta5}
the wavefunction of the $f$-electron state for $\lambda/U \sim 0.1$ is well
approximated by that in the $j$-$j$ coupling scheme.
The $f$-electron state is continuously changed from the $LS$ to
the $j$-$j$ coupling limits, when we change the ratio of $\lambda/U$,
but the transition region is found in the range between $\lambda/U=0.1$ and $1$.

For $f^7$-electron case, we also find such a transition in
the same parameter region of $\lambda/U$.
In order to reconfirm this point for the case of $n=7$,
we evaluate the Curie constant $C$ on the basis of $H_{\rm loc}$
for $W=0$.
We obtain the magnetic susceptibility $\chi$ as
$\chi=C/T$ and the Curie constant $C$ is given by
$C=(g_J \mu_B)^2 J(J+1)/3k_{\rm B}$.
Here $\mu_{\rm B}$ indicates the Bohr's magneton,
$J$ denotes the size of total angular momentum ${\mib J}$,
which is given by ${\mib J}={\mib L}+{\mib S}$
with total angular momentum ${\mib L}$ and total spin momentum ${\mib S}$
and $g_J$ is the Land\'e's $g$-factor.
Note that this expression for $C$ is in common with the $LS$ and $j$-$j$ coupling limits,
although the magnitude of $C$ is changed.

\begin{figure}[t]
\centering
\includegraphics[width=8.0truecm]{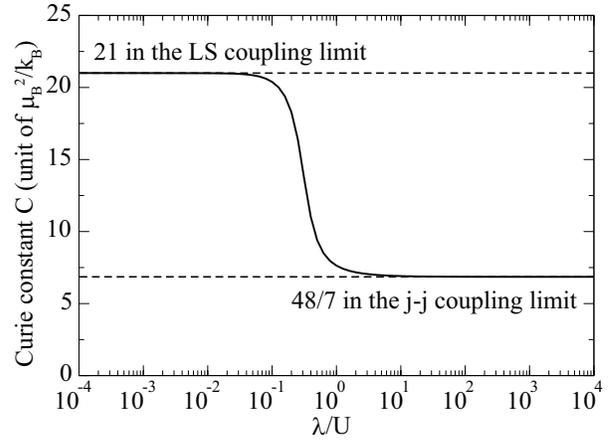}
\caption{
Curie constant $C$ for $n=7$ as a function of $\lambda/U$ for the case of $W=0$.
We find the transition from the value for the $LS$ coupling limit to that for the $j$-$j$ coupling limit
in the region pf $0.1 < \lambda/U < 1$, which includes the parameters for actual materials.
}
\end{figure}

In the $LS$ coupling limit, for the case of $n=7$,
we obtain $J=S=7/2$ and $L=0$ due to the Hund's rules.
In this case, we find $g_J=2$, which is just equal to
the electron's $g$ factor.
Thus, we obtain the magnetic moment as $M=7 \mu_{\rm B}$
and $C=21 \mu_{\rm B}^2/k_{\rm B}$.
On the other hand, in the $j$-$j$ coupling limit,
we accommodate one $f$ electron in the octet $j=7/2$,
while the sextet $j=5/2$ is fully occupied.
In this case, the total angular momentum $J$ is, of course, $7/2$,
which is originating from one $f$ electron in $j=7/2$ octet.
The Land\'e' $g$-factor of $j=7/2$ is equal to $g_J=8/7$
and thus, we obtain $M=4 \mu_{\rm B}$ and
$C=(48/7)\mu_{\rm B}^2/k_{\rm B}$.
Due to the reduction of the magnetization,
 $C$ is changed from $21$ to $48/7$.

Concerning the value of $C$ between the $LS$ and $j$-$j$ coupling limits,
it is necessary to perform the numerical evaluation of
eigenenergies and eigenstates of $H_{\rm loc}$ for
$n=7$ and $W=0$ by changing the ratio of $\lambda/U$.
The result is summarized in Fig.~2.
As mentioned above, we find $C=21$ and $48/7$ in the limit
of $\lambda/U=0$ and $\lambda/U=\infty$, respectively,
leading to the good check of the numerical calculations.
We note the continuous decrease in $C$ from $21$ to $48/7$,
which originates from the reduction of the
magnetic moment $M$ from $M=7 \mu_{\rm B}$
in the $LS$ coupling limit to $M=4 \mu_{\rm B}$
in the $j$-$j$ coupling limit.

It is emphasized here that the reduction of $C$ occurs
in the range of $0.1 < \lambda/U < 1$,
which just corresponds to the values of actual $f$-electron materials.
In this transition region, the $f$-electron wavefunction is given
by the mixture of those in the $LS$ and $j$-$j$ coupling limits.
Thus, the properties of both limits can be observed
in the range of $0.1 < \lambda/U < 1$,
leading to the expectation of a possible explanation of
the Kondo behavior in $f^7$ electron systems.

\begin{figure}[t]
\centering
\includegraphics[width=8.0truecm]{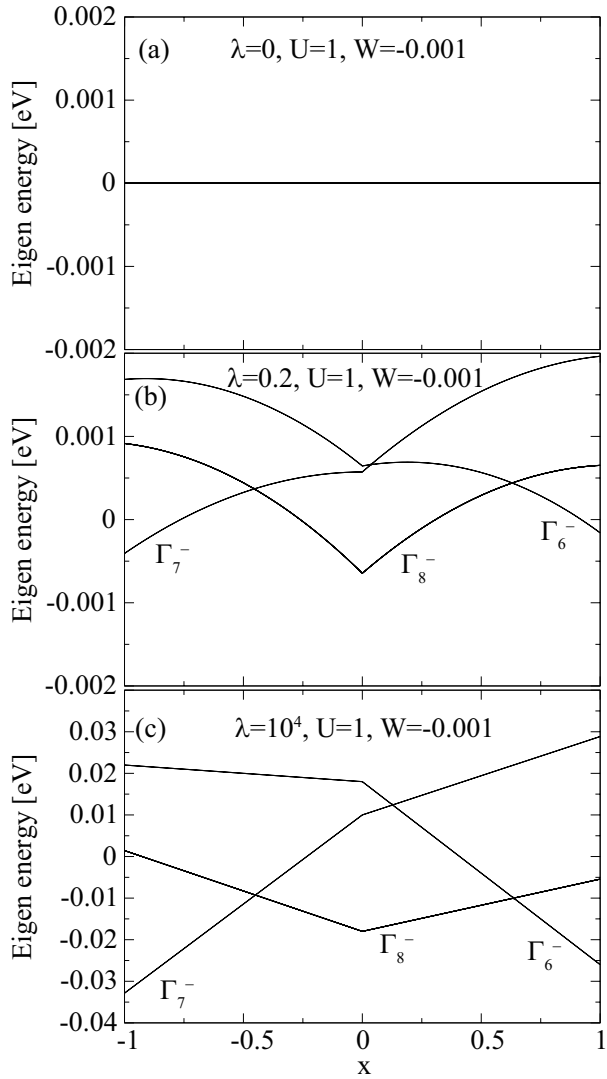}
\caption{(a) Eigenenergies vs. $x$ for $\lambda=0$, $U=1$, and $W=-10^{-3}$.
(b) Eigenenergies vs. $x$ for $\lambda=0.2$, $U=1$, and $W=-10^{-3}$.
(c) Eigenenergies vs. $x$ for $\lambda=10^{4}$, $U=1$, and $W=-10^{-3}$.
Note that (a) denotes the result in the LS coupling scheme, while (c) indicates
that in the $j$-$j$ coupling limit.
}
\end{figure}

\subsection{CEF level schemes}

Another evidence for the mixture of both limits is
found in the CEF level schemes.
In Figs.~3, we summarize the variation of the CEF level schemes,
when $\lambda$ is increased for the fixed values of $U=1$
and $W=-10^{-3}$.
Note that Fig.~3(a) indicates the result in the $LS$ coupling limit,
in which we do not find any effects of CEF potentials on the $f$-electron
low-energy states for $n=7$.
Readers may be doubtful of this result, but we should note that
$J$ is purely given by spin, i.e.,
$J=S=7/2$ in the $LS$ coupling limit for $n=7$.
When we recall the fact that the CEF potential acts on charge,
not on spin, there occurs no effect of CEF potentials on the states
with $J=S=7/2$, except for the shift of the total energy.
Thus, we still obtain the octet independent of CEF potentials.

Next we consider the $j$-$j$ coupling limit.
Here we pay our attention to Fig.~3(c) by skipping Fig.~3(b).
In Fig.~3(c), we exhibit the result for $\lambda=10^{4}$.
This is, of course, unphysical value, but we use it
for the purpose to realize the $j$-$j$ coupling limit.
As mentioned above, in this limit, we accommodate six electrons
in the sextet of $j=5/2$ and one electron in the octet of $j=7/2$.
In sharp contrast to the $LS$ coupling limit with $J=S=7/2$,
one $f$ electron in the $j=7/2$ octet feels the CEF potentials,
as shown in Fig.~3(c).
Under the cubic CEF potentials, the octet is split into
two doublets ($\Gamma_6^{-}$ and $\Gamma_7^{-}$)
and one quartet ($\Gamma_8^{-}$).
It is quite natural that the present level scheme is quite similar
to that for $J=7/2$~\cite{LLW} due to the symmetry reason,
although there are small deviations due to the difference
in the definition of the CEF parameters.

Let us turn our attention back to Fig.~3(b).
This is the result for $\lambda=0.2$, which is near
the realistic values for Eu and Gd ions.
First we notice that the energy scale of Fig.~3(b)
is smaller in one order than that of Fig.~3(c).
The CEF energy splitting is in the order of meV,
which is about 10 K.
This value is considered to be small, but the CEF excitation
in this order can be detected in the experiment.
Thus, as mentioned in the previous paper,\cite{Hotta1}
it is a challenging issue to measure the CEF excitation
in Gd and Eu compounds, which have not been expected to
show CEF excitations.

Second the symmetries of the CEF ground states are the same as those
in the $j$-$j$ coupling scheme.
It is quite natural that the ground-state multiplet is always characterized
by the total angular momentum $J=7/2$, irrespective of the values
of $U$ and $\lambda$.
Note that the magnitude of the matrix elements among the states
of $J_z$ is affected by $U$ and $\lambda$, where $J_z$ denotes the
$z$-component of  $J$.

\section{Model and Method}

\subsection{Seven-orbital Anderson model}

Now we consider the conduction electron hybridized with localized electrons.
In general, all $f$ orbitals are hybridized with conduction electrons with
the same symmetry, but it is very difficult to perform numerical calculations
by taking into account seven conduction bands.
Since the purpose here is to reveal the effect of spin-orbit coupling
on the Kondo phenomena in $f^7$ electron systems,
it is reasonable to consider the minimum model to investigate the Kondo phenomena.
In this sense, the minimum model should include one conduction band,
since we can expect the appearance of the underscreening Kondo phenomena
even for one conduction band hybridized with local $f$ orbitals.
Then, it is natural to consider $a_{\rm u}$ conduction band,
since local $a_{\rm u}$ state is non-degenerate even under the high symmetry
ligand field such as the cubic CEF potential.
Note that the local $a_{\rm u}$ state is described as
$(f^{\dag}_{m=2,\sigma}-f^{\dag}_{m=-2,\sigma})|0\rangle /\sqrt{2}$,
where $|0\rangle$ denotes the vacuum.

Then, the model is expressed as
\begin{equation}
  H=\sum_{\mib{k},\sigma} \varepsilon_{\mib{k}}
    c_{\mib{k}\sigma}^{\dag} c_{\mib{k}\sigma}
   +\sum_{\mib{k},m,\sigma} (V_m c_{\mib{k}\sigma}^{\dag}f_{m\sigma}+{\rm h.c.})
   +H_{\rm loc},
\end{equation}
where $\varepsilon_{\mib{k}}$ is the dispersion of conduction electron,
$c_{\mib{k}\sigma}$ is an annihilation operator of conduction electron
with momentum $\mib{k}$ and spin $\sigma$,
and $V_m$ is the hybridization between conduction and localized electrons.
Note that $V_{m}$ is given by $V_{m=2}=-V_{m=-2}=V$ and zero
for other components.
The energy unit is a half of the conduction bandwidth,
which is set as 1 eV throughout this paper, as mentioned above.

\subsection{Numerical renormalization group method}

For the diagonalization of the impurity Anderson model, we employ
a numerical renormalization group (NRG) method,\cite{NRG1,NRG2}
in which we  logarithmically discretize the momentum space so as
to include efficiently the conduction electrons near the Fermi energy.
The conduction electron states are characterized by ``shell'' labeled by $N$
and the shell of $N=0$ denotes an impurity site described by the local Hamiltonian.
Then, after some algebraic calculations,
the Hamiltonian is transformed into the recursion form as
\begin{eqnarray}
  H_{N+1} = \sqrt{\Lambda} H_N+\xi_N \sum_{\sigma}
  (c_{N\sigma}^{\dag}c_{N+1\sigma}+c_{N+1\sigma}^{\dag}c_{N\sigma}),
\end{eqnarray}
where $\Lambda$ is a parameter for logarithmic discretization,
$c_{N\sigma}$ denotes the annihilation operator of conduction electron
in the $N$-shell, and $\xi_N$ indicates ``hopping'' of electron between
$N$- and $(N+1)$-shells, expressed by
\begin{eqnarray}
  \xi_N=\frac{(1+\Lambda^{-1})(1-\Lambda^{-N-1})}
  {2\sqrt{(1-\Lambda^{-2N-1})(1-\Lambda^{-2N-3})}}.
\end{eqnarray}
The initial term $H_0$ is given by
\begin{equation}
  H_0=\Lambda^{-1/2} [H_{\rm loc}
  +\sum_{m\sigma}V_m(c_{0\sigma}^{\dag}f_{m\sigma}
                             +f_{m\sigma}^{\dag}c_{0\sigma})].
\end{equation}

For the calculations of thermodynamic quantities,
We evaluate the free energy $F$ for local $f$ electron in each step by
\begin{eqnarray}
   F = -T (\ln {\rm Tr} e^{-H_N/T} - \ln {\rm Tr} e^{-H_N^0/T}),
\end{eqnarray}
where a temperature $T$ is defined as $T=\Lambda^{-(N-1)/2}$
in the NRG calculation and $H_N^0$ denotes the Hamiltonian
without the hybridization term and $H_{\rm loc}$.
Then, we obtain the entropy $S_{\rm imp}$ by
$S_{\rm imp}=-\partial F/\partial T$
and the specific heat $C_{\rm imp}$ is evaluated by
$C_{\rm imp}=-T\partial^2 F/\partial T^2$.
In the NRG calculation, we keep $M$ low-energy states
for each renormalization step.
In this paper, we set $\Lambda=5$ and we keep $M=4,500$
low-energy states for each renormalization step.

\subsection{Multipole susceptibility}

For the purpose to discuss the multipole properties later,
we provide the definition of multipole operator in this subsection.
Since we discuss the effect of the spin-orbit coupling
for $\lambda/U=0$ to $\lambda/U=\infty$,
it is necessary to use the same definition both in the $LS$
and $j$-$j$ coupling schemes, irrespective of the values of $\lambda/U$.
Thus, we define the multipole as a spin-orbital density in the form of
a one-body operator from the viewpoint of
the multipole expansion of electron density in electromagnetism.
On the basis of this definition of the multipole operator,
we have developed microscopic theories
for multipole-related phenomena.
For instance, octupole ordering in NpO$_2$ has been clarified
by evaluating multipole interaction
by the standard perturbation method
in terms of electron hopping.\cite{Kubo1,Kubo2,Kubo3}
We have discussed possible multipole states of
filled skutterudites by analyzing the multipole susceptibility of
a multiorbital Anderson model based on the $j$-$j$ coupling scheme.
\cite{Hotta1,Hotta2,Hotta3,Hotta4,Hotta5,Hotta6,Hotta7,Hotta8}
We have also discussed the multipole state in actinide dioxides~\cite{Hotta9,Hotta10}
and Yb compounds.\cite{Hotta11}
Recently, a microscopic theory for multipole ordering from an itinerant picture
has been developed on the basis of a seven-orbital
Hubbard model with spin-orbit coupling.\cite{Hotta12}

The multipole operator ${\hat T}$ is expressed
in the second-quantized form as
\begin{equation}
  {\hat T}^{(k)}_{\mib{i},\gamma} = \sum_{q,m\sigma,m'\sigma'}
  G^{(k)}_{\gamma,q}
  T^{(k,q)}_{m\sigma,m'\sigma'}f^{\dag}_{\mib{i}m\sigma}f_{\mib{i}m'\sigma'},
\end{equation}
where $k$ indicates the rank of the multipole,
$q$ denotes an integer between $-k$ and $k$,
$\gamma$ is a label used to express an $O_{\rm h}$ irreducible representation,
$G^{(k)}_{\gamma,q}$ is the transformation matrix
between spherical and cubic harmonics,
and $T^{(k,q)}_{m\sigma,m\sigma'}$ can be calculated,
using the Wigner-Eckart theorem, as \cite{Inui}
\begin{equation}
 \label{Tkq}
 \begin{split}
  T^{(k,q)}_{m\sigma,m'\sigma'}
  &= \sum_{j,\mu,\mu'}
  \frac{\langle j || T^{(k)} || j \rangle}{\sqrt{2j+1}}
  \langle j \mu | j \mu' k q \rangle \\
  &\times 
  \langle j \mu | \ell m s \frac{\sigma}{2} \rangle
  \langle j \mu' | \ell m' s \frac{\sigma'}{2} \rangle.
 \end{split}
\end{equation}
Here, $\ell=3$, $s=1/2$, $j=\ell \pm s$,
$\mu$ denotes the $z$-component of $j$,
$\langle j \mu | j' \mu' j'' \mu'' \rangle$ indicates
the Clebsch-Gordan coefficient,
and $\langle j || T^{(k)} || j \rangle$ is
the reduced matrix element for a spherical tensor operator
and is given by
\begin{equation}
  \label{red}
  \langle j || T^{(k)} || j \rangle=
  \frac{1}{2^k} \sqrt{\frac{(2j+k+1)!}{(2j-k)!}}.
\end{equation}
Note that $k \le 2j$ and the highest rank is $2j$.
Thus, we treat multipoles up to rank 7 for $f$ electrons
in this definition.

Here we should note that multipoles belonging to the same symmetry
are mixed in general, even if the rank is different.
Namely, the $f$-electron spin-charge density should be given by
the appropriate superposition of multipoles, expressed as
\begin{equation}
  \label{multi}
  {\hat X}=\sum_{k,\gamma} p^{(k)}_{\gamma}{\hat T}^{(k)}_{\gamma},
\end{equation}
where we redefine each multipole operator so as to satisfy
the orthonormal condition of~\cite{Kubo3}
\begin{equation}
  \label{normal}
  {\rm Tr} \Bigl\{
  {\hat T}_{\gamma}^{(k)} {\hat T}_{\gamma'}^{(k')} \Big\}
  =\delta_{kk'}\delta_{\gamma\gamma'}.
\end{equation}
In order to determine the coefficient $p^{(k)}_{\gamma}$, it is necessary
to evaluate the multipole susceptibility in the linear response theory.
Namely, $p^{(k)}_{\gamma}$ is determined by the eigenstate of the largest
eigenvalue of susceptibility matrix, given by
\begin{equation}
 \label{sus}
  \begin{split}
   \chi_{k\gamma,k'\gamma'} \!
   =\! & \frac{1}{Z}
   \sum_{i,j} \frac{e^{-E_i/T}-e^{-E_j/T}}{E_j-E_i}
   \langle i | [{\hat T}^{(k)}_{\gamma}-\rho^{(k)}_{\gamma}] | j \rangle
   \\ & \times 
   \langle j | [{\hat T}^{(k')}_{\gamma'}- \rho^{(k')}_{\gamma'}]| i \rangle,
 \end{split}
\end{equation}
where $Z$ is the partition function given by $Z$=$\sum_i e^{-E_i/T}$,
$E_i$ denotes the eigenenergy of the $i$-th eigenstate
$|i\rangle$ of the Hamiltonian, $T$ is a temperature, and
$\rho^{(k)}_{\gamma}$=$\sum_i e^{-E_i/T}
\langle i |{\hat T}^{(k)}_{\gamma}| i \rangle/Z$.
The multipole susceptibility is given by
the eigenvalue of the susceptibility matrix.
By using the NRG method, we evaluate the matrix elements of
the multipole susceptibility in each NRG step.
Then, we can determined the optimized multipole state
in an unbiased manner.

Note that in this paper, we express the irreducible
representation of the CEF state by Bethe notation,
whereas for multipoles, we use short-hand notations by the
combination of the number of irreducible representation and
the parity of time reversal symmetry, ``g'' for gerade and ``u'' for
ungerade.
For O$_{\rm h}$ symmetry, we have nine independent
multipole components as 1g, 2g, 2u,
3g, 3u, 4g, 4u, 5g, and 5u.
Note that 1u does not appear within the rank 7.

\section{Calculation Results}

\subsection{Underscreening Kondo effect}

Let us show our numerical results.
First we consider the situation of $U \ne 0$ and $\lambda=W=0$,
which is the limit of $LS$ coupling scheme.
In this case, the ground-state multiplet is characterized by $S=7/2$
and $L=0$.
When one conduction band is hybridized with such localized state,
it has been well known that the underscreening Kondo effect occurs.
Namely, one of seven electron among the octet forms the singlet due to the hybridization
with single conduction electron band, while another six electrons are still localized.
Thus, even after the underscreening Kondo effect, the septet remains.
Here we use the Kondo temperature, even when the underscreening
Kondo effect occurs.

\begin{figure}[t]
\centering
\includegraphics[width=8.0truecm]{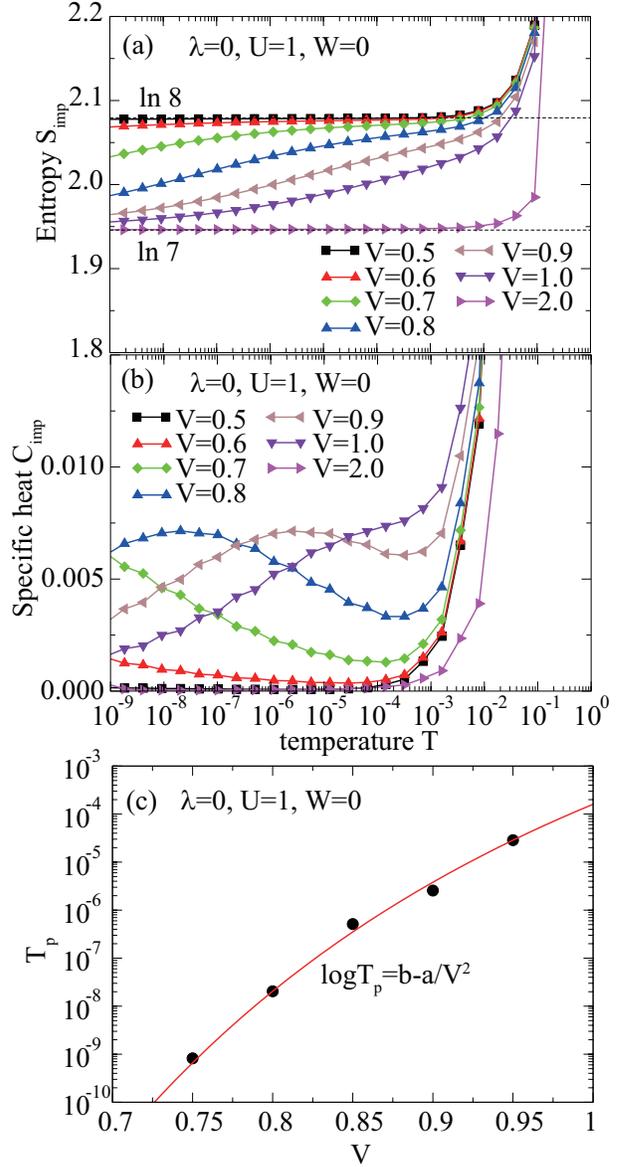}
\caption{(Color online)
(a) Entropy $S_{\rm imp}$ vs. temperature $T$ for $U=1$ and $\lambda=W=0$.
The hybridization $V$ is changed between $0.5$ and $2.0$.
(b) Specific heat $C_{\rm imp}$ vs. temperature for the same parameters in (a).
(c) The peak temperature $T_{\rm p}$ (solid symbols) vs. $V$.
The red curve denotes $\ln T_{\rm p}=b-a/V^2$ with appropriate constants $a$ and $b$.
}
\end{figure}

In Fig.~4(a), we show the temperature dependence of entropy
for several values of $V$.
For $V$=0.5, we find the entropy $\ln 8$ in the present temperature range.
If we decrease the temperature, we expect the change of the entropy, but
it is difficult to obtain the results with a reliable precision.
Then, we increase the value of $V$ to elevate the Kondo temperature.
As expected, we find the gradual decrease in the entropy when $V$ is increased.
For $V$=2.0, the entropy immediately becomes $\ln 7$ in the present
temperature range, since the Kondo temperature is high enough.

Although it is difficult to define the Kondo temperature only from Fig.~4(a),
it is possible to define $T_{\rm K}$ as a peak temperature in the
specific heat, which is formed by the entropy change from $\ln 8$ to $\ln 7$.
The results are shown in Fig.~4(b).
Note that the absolute values of the specific heat are suppressed,
since the entropy release is relatively small in this case.
Tiny fluctuations in the specific heat are observed,
since the magnitude of the specific heat is relatively small in this case and
the effect of numerical precision appears.
For $V=0.5$, we find $C_{\rm imp}$ is zero and no peak is found in the region
of $T<10^{-3}$.
When $V$ is increased, the value of $C_{\rm imp}$ is gradually enhanced.
For $V=0.8$ and $0.9$, we find the broad peak structure in the specific heat.
These peaks are considered to originate from the entropy change
from $\ln 8$ to $\ln 7$.
For $V=1.0$, we observe a weak shoulder structure around at $T=10^{-4}$,
but the peak structure is smeared.
For $V=2.0$, $C_{\rm imp}$ becomes immediately zero and no peak structure
can be found for $T<10^{-3}$.

In oder to confirm that the entropy change and the peak formation in the specific heat
are indications of the Kondo effect, we plot the peak temperature $T_{\rm p}$ in Fig.~4(c).
We can define the peaks in the region of $0.75 < V < 1$, but we find that $T_{\rm p}$
is well fitted by $b{\rm exp}(-a/V^2)$ with appropriate constants $a$ and $b$.
This is expected from the well-known formula
for the Kondo temperature $T_{\rm K}=D{\rm exp}[-1/(\rho J_{\rm cf})]$,
where $D$ is the half of the conduction electron band.
Note that $D$ is set as unity in this paper.
In the single-band Anderson model, we can obtain $J_{\rm cf}=4V^2/I$,
where $I$ denotes the on-site Coulomb interaction.
In the present case, it is difficult to derive such a simple form of $J_{\rm cf}$,
since we consider the complicated orbital-dependent interactions through
the Slater-Condon parameters.
However, we can deduce that $J_{\rm cf}$ can be obtained, in any case,
in the second-order perturbation in terms of $V$.
We expect that $T_{\rm p}$ is in proportion to ${\rm exp}(-a/V^2)$
with an appropriate constant $a$.
We also introduce another constant $b$ to fit $T_{\rm p}$ as $b{\rm exp}(-a/V^2)$.
From the result in Fig.~4(c), we conclude that the underscreening Kondo effect
actually occurs and it is characterized by the entropy change from $\ln 8$ to $\ln 7$
due to the screening of one spin by single conduction band.

\subsection{Effect of spin-orbit coupling}

Next we include the spin-orbit coupling.
Namely, we consider the finite values of $\lambda$ for $U=1$ and $W=0$.
Depending on the values of $V$, we consider two typical situations
for large and small values of $V$ such as $V=3.0$ and $0.6$.
From Fig.~4(a), the underscreening Kondo effect occurs in the
high-temperature region as $T>0.1$ for $V>2.0$,
while the octet still remains in the temperature range for $V<0.6$.

\begin{figure}[t]
\centering
\includegraphics[width=8.0truecm]{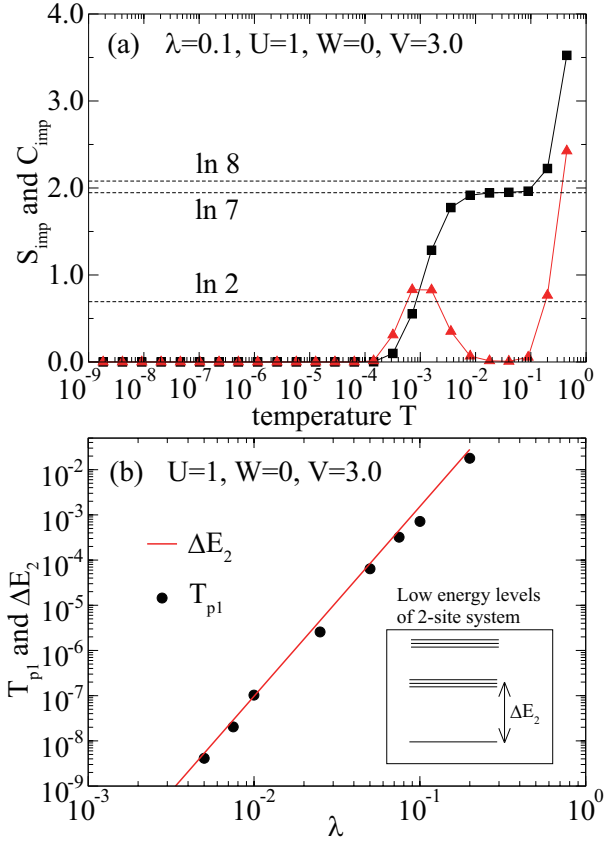}
\caption{(Color online)
(a) Entropy $S_{\rm imp}$ (solid square) and specific heat $C_{\rm imp}$
(solid triangle) for $\lambda=0.1$, $U=1$, $W=0$, and $V=3.0$.
(b) The peak temperature $T_{\rm p1}$ (solid circle) and the excitation energy
$\Delta E_2$ (solid line) of the two-site system as functions of $\lambda$.
Inset shows the schematic view for the energy levels of the two-site system,
in which the septet is split into one singlet and two triplets.
}
\end{figure}

In Fig.~5(a), we show the typical results for entropy and specific heat
for $\lambda=0.1$,  $U=1$, and $W=0$ with the large value of $V$
such as $V=3.0$.
As expected from the results in Figs.~4, we observe the plateau of $\ln 7$
for $T<0.1$.
When we further decrease the temperature, we find the release of entropy
$\ln 7$, leading to a large peak in the specific heat, as shown in Fig.~5(a).
In the present case, we find such a peak at $T \approx 10^{-3}$.
Here we define the peak temperature as $T_{\rm p1}$.
We are interested in the origin of this entropy change, but
it seems to be irrelevant to the Kondo effect.
Intuitively, the entropy change in this case is rather rapid
in comparison with the that in the Kondo effect.
A more concrete discussion can be done by the direct comparison
of the relevant energy scales.
In one word, this is the level splitting due to the hybridization.

At present we do not explicitly include the CEF potential, but
it has been well known that the level splitting occurs
due to the effect of the hybridization between localized and
conduction electrons.
In fact, as shown in Fig.~5(b), the peak temperature $T_{\rm p1}$
is well fitted by $\Delta E_2$, where $\Delta E_2$ denotes the
excitation energy between the ground state singlet and the first excited
triplet in a two-site system, as shown in the inset of Fig.~5(b).
Note that the two-site system is composed of impurity site and one conduction site,
which are connected by the hybridization $V$.
For the case of $\lambda=0$, we find that the ground state is septet.
Due to the combination of $S=7/2$ at impurity site and $S=1/2$ at the conduction site,
we obtain $S_{\rm tot}=4$ and $3$, but the state with $S_{\rm tot}=3$
becomes the ground state with septet.
Here $S_{\rm tot}$ denotes the magnitude of total spin moment.
When we include the effect of $\lambda$, the level splitting due to the hybridization
becomes significant and $\Delta E_2$ is increased with the increase of $\lambda$.
The temperature to characterize the entropy release of $\ln 7$ is deduced to
be determined by $\Delta E_2$, as shown in Fig.~5(b).

\begin{figure}[t]
\centering
\includegraphics[width=8.0truecm]{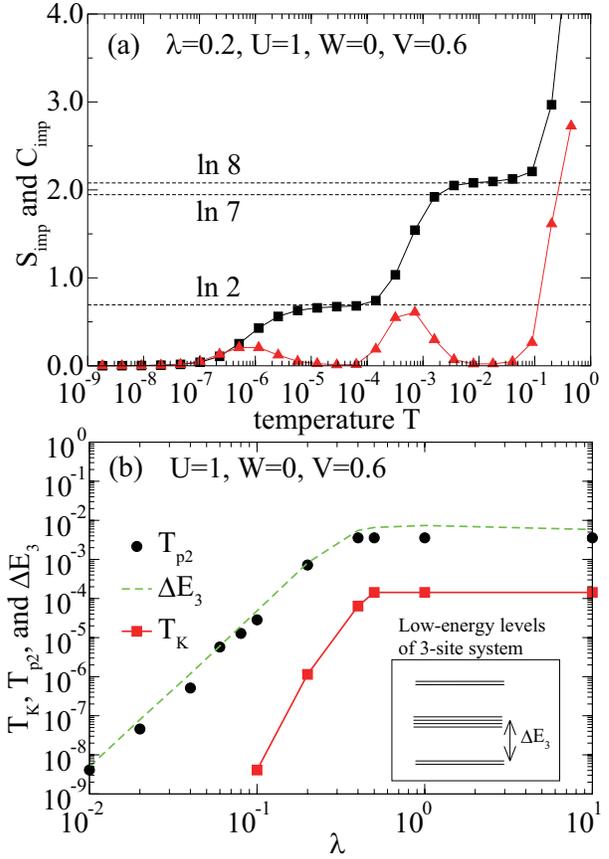}
\caption{(Color online)
(a) Entropy $S_{\rm imp}$ (solid square) and specific heat $C_{\rm imp}$
for $\lambda=0.2$, $U=1$, $W=0$, and $V=0.6$.
(b) The peak temperature $T_{\rm p2}$ (solid circle), the Kondo temperature $T_{\rm K}$
(solid square with solid line), and the excitation energy $\Delta E_3$ (broken curve)
of the three-site system as functions of $\lambda$.
Inset shows the schematic view of the energy levels of the three-site system,
in which the octet is split into one quartet and two doublets.
}
\end{figure}

Next we consider the case with small $V$.
In Fig.~6(a),  we depict the entropy and specific heat
for $\lambda=0.2$, $U=1$, $W=0$, and $V=0.6$.
As expected from the discussion in Figs.~5, at high temperatures
as $10^{-3}<T<10^{-1}$, we find a plateau of $\ln 8$
originating from the local octet of $J=7/2$.
For the case of $\lambda=0$, only one electron spin is screened by
one conduction band, leading to the entropy change of $\ln 8$ to
$\ln 7$, but in the present case with $\lambda=0.2$, first we find
the entropy change from $\ln 8$ to $\ln 2$ by the level splitting
due to the hybridization with conduction electrons.
Then, the remained $\ln 2$ from the local doublet is eventually
released by the Kondo effect.

We claim that the result in Fig.~6(a) is important.
The underscreening Kondo effect in the $f^7$-electron system
has been well understood and probably, it has not been considered to
be an intriguing phenomenon.
However, under the effect of the spin-orbit coupling,
the entropy of $\ln 8$ from the local octet due to $J=S=7/2$ is
released in the two step.
Thus, in the specific heat, we observe two peak temperatures.
The higher and lower ones are defined as 
$T_{\rm p2}$ and $T_{\rm K}$, respectively.

In Fig.~6(b), we plot  $T_{\rm p2}$ and $T_{\rm K}$ as functions of
$\lambda$.
Both temperatures are increased monotonically up to $\lambda=0.5$,
but for $\lambda>0.5$, those seem to be almost constant,
even if we change $\lambda$.
The higher peak temperature $T_{\rm p2}$ is found to be
well fitted by $\Delta E_3$, which is the excitation energy
in the three-site system, including one impurity and two conduction sites.
The ground state of the three-site system is found to be characterized
by $S_{\rm tot}=7/2$ for the case of $\lambda=0$.
When we include the spin-orbit coupling, the octet ground state
is split into three, two doublets and one quartet.
Among them, one doublet becomes the ground state and the quartet
is the first excited state.
The energy difference between them is here defined as $\Delta E_3$.
Again this is the level splitting due to the effect of hybridization.
Note that the $\lambda$ dependence of $\Delta E_2$ is similar to
that of $\Delta E_3$, but the magnitude is found to be different
from $T_{\rm p2}$.

As we observed in Fig.~6(b), the higher peak $T_{\rm p2}$ is
scaled by $\Delta E_3$.
The lower peak is considered to be the Kondo temperature $T_{\rm K}$.
We find that $T_{\rm K}$ is increased monotonically up to $\lambda=0.5$
and it becomes constant for $\lambda > 0.5$.
In the limit of $\lambda=\infty$, the $j$-$j$ coupling scheme becomes
exact and we know that one electron is accommodated in $j=7/2$ octet.
In such a limit, as emphasized in the section of introduction,
the hybridization of one $j=7/2$ electron with the conduction band
leads to the Kondo effect.
For enough large $\lambda$, we easily deduce that the situation is not
so changed from that in the limit of $\lambda=\infty$ and thus,
it seems natural that the present $T_{\rm K}$ is constant
in the region of large $\lambda$.

However, it is surprising that the same $T_{\rm K}$ as in the $j$-$j$ coupling
limit is obtained even for $\lambda=0.5$, which is not considered to
be large enough.
When we decrease $\lambda$, $T_{\rm K}$ is decreased rapidly,
but we still find $T_{\rm K}$ in the present temperature range
for $\lambda=0.1$.
Thus, we arrive at an important conclusion that the picture of the
Kondo effect on the basis of the $j$-$j$ coupling scheme is
applicable even for the realistic situation with the spin-orbit coupling
in the order of $\lambda/U=0.1$ for $f^7$ electron systems.
The component of the $j$-$j$ coupling scheme in the many-electron
wave function is more persevering than we have naively expected.

\begin{figure}[t]
\centering
\includegraphics[width=8.0truecm]{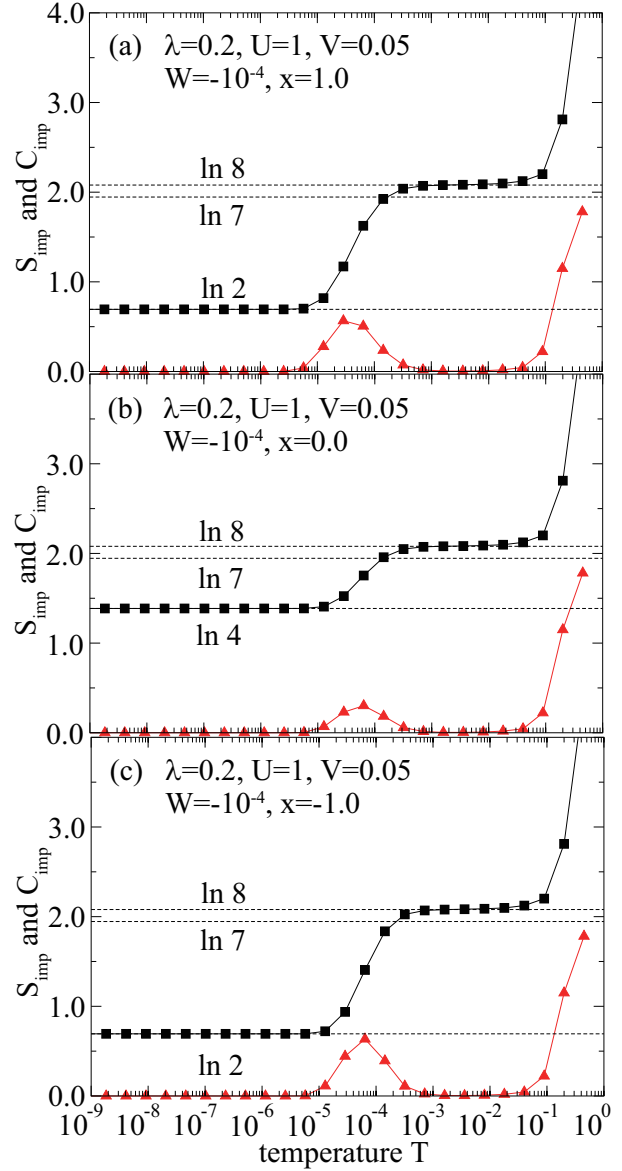}
\caption{(Color online)
Entropy $S_{\rm imp}$ (solid square) and specific heat $C_{\rm imp}$ (solid triangle)
for (a) $x=1.0$, (b) $x=0.0$, and (c) $x=-1.0$.
Other parameters are set as $\lambda=0.2$, $U=1$, $W=-10^{-4}$, and $V=0.05$.
}
\end{figure}

\subsection{Effect of CEF potential}

Thus far, we have considered the situation without the CEF potential.
As mentioned above, due to the effect of the hybridization,
the level splitting has been found to occur.
Next we include explicitly the CEF potentials.
In Figs.~7, we summarize the results of entropy and specific heat
for three CEF ground states, which are controlled by $x$.
As easily understood from Fig.~3(b),
the cases of $x=1.0$, $0.0$, and $-1.0$ correspond to the ground states of
$\Gamma_6^{-}$, $\Gamma_8^{-}$, and $\Gamma_7^{-}$, respectively.
Since $V$ is chosen to be very small here, we find the localized CEF states in
the present temperature range.
The residual entropy is $\ln 4$ for $\Gamma_8^-$ quartet,
while it is $\ln 2$ for $\Gamma_6^-$ or $\Gamma_7^-$ doublet.

The entropy release from $\ln 8$ to $\ln 4$ or $\ln 2$ occurs
and we find the peak in the specific heat at the corresponding temperature.
The height of the peak depends on the released entropy, but
the position of the peak is almost the same for three cases.
The peak position is around $10^{-4}$ and this scale is determined
by the CEF level splitting.
As understood from Fig.~3(b), for $\Lambda=0.2$, $U=1$, and $W=-0.001$,
the order of the CEF level splitting is in the order of $10^{-3}$.
In the present calculation, we set $W=-10^{-4}$, and thus,
the order of the CEF level splitting is smaller in one order.
In this sense, it is quite natural that the peak position appears
around at $T=10^{-4}$ in common among Figs.~7(a)-7(c).

\begin{figure}[t]
\centering
\includegraphics[width=8.0truecm]{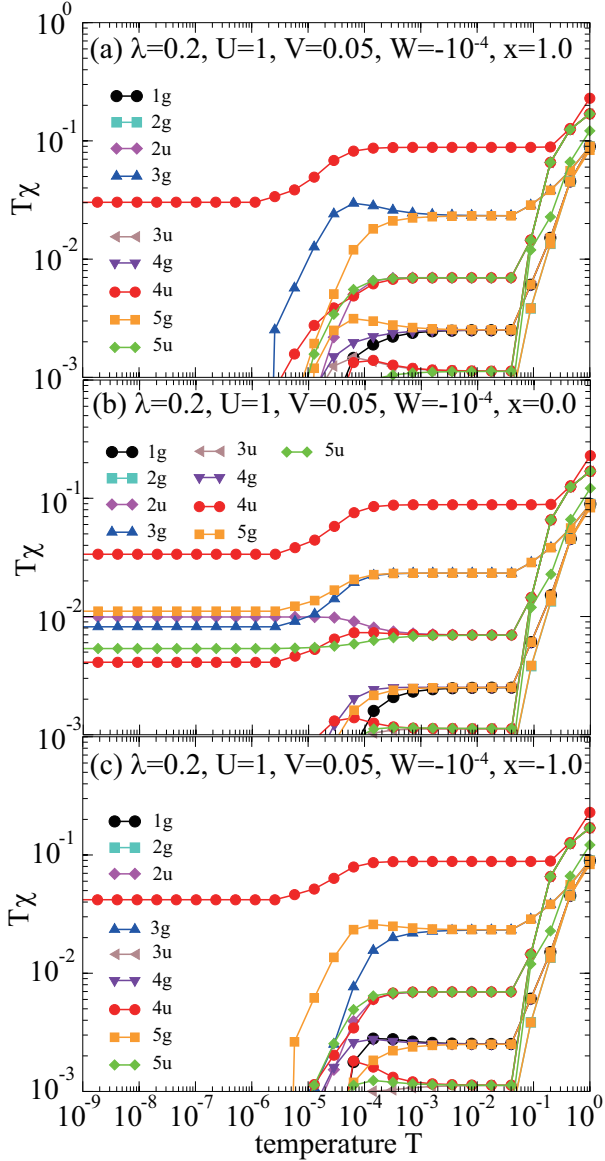}
\caption{(Color online)
Multipole susceptibilities for (a) $x=1.0$, (b) $x=0.0$, and (c) $x=-1.0$.
Other parameters are set as $\lambda=0.2$, $U=1$, $W=-10^{-4}$, and $V=0.05$.
}
\end{figure}

Next we consider the multipole properties.
After we faithfully diagonalize the multipole susceptibility matrix eq.~(\ref{sus}),
we obtain the eigenvalues and eigenvectors.
The multipole state with the largest eigenvalue is considered to
be realized and the corresponding eigenvectors denote the
optimized multipole state given by the mixture of multipole
components with different ranks in the same symmetry group.

First we consider the case without the CEF potential, i.e., $W=0$.
The results are not shown here, but the optimized multipole is
always dipole for $\lambda=0$, since $J$ is purely equal to
total spin moment $S=7/2$.
When we increase the value of $\lambda$, we observe the
increase of other multipole components and in the $j$-$j$ coupling limit,
one electron in the $j$=7/2 octet carries all multipoles higher than dipole.
Note that in a periodic system, the optimized multipole is affected by
the lattice structure, electron hopping amplitude, and interactions.

Next we include the CEF potential.
In Figs.~8, we show the eigenvalues of multipole susceptibility matrix
in each NRG step for the same parameters in Figs.~7.
When we see the entropy and specific heat, it is difficult to
find the difference in the cases for
$\Gamma_6^-$ and $\Gamma_7^-$ ground states,
corresponding to Figs.~7(a) and 7(c), respectively.
In the plateau of $\ln 8$, there are no significant differences
among three cases, but around at the temperature of the entropy release,
it is possible to detect the different multipole states for three cases.
For the case of $\Gamma_8^-$ ground state,
since $\Gamma_8^-$ includes orbital degrees of freedom,
it carries higher-rank multipoles.
On the other hand, $\Gamma_6^-$ and $\Gamma_7^-$ states
carry only 4u multipoles, mainly characterized by dipole.

However, for $\Gamma_6^-$ and $\Gamma_7^-$ ground states,
we can observe the difference in the multipole state with
the second largest eigenvalue.
Namely, they are 3g quadrupole and 5g quadrupole, respectively,
for Figs.~8(a) and 8(c).
The multipole with the largest eigenvalue is always 4u dipole,
but the Curie constant for the multipole susceptibility is slightly different.
Namely, in the $j$-$j$ coupling limit, we obtain the dipole moment
as $(4/3) \mu_{\rm B}$ and $(12/7) \mu_{\rm B}$ for $\Gamma_6^-$
and $\Gamma_7^-$ ground states, respectively.
This difference appears in the values of $T\chi$ at the lowest
temperatures in the present calculations.

\begin{figure}[t]
\centering
\includegraphics[width=8.0truecm]{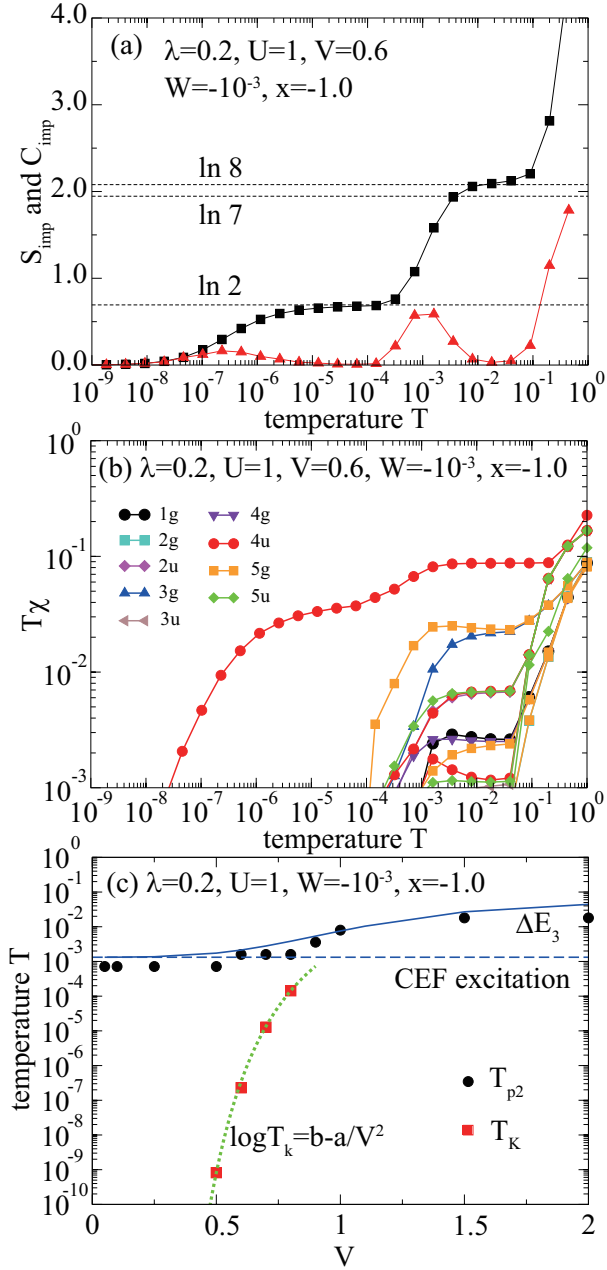}
\caption{(Color online)
(a) Entropy $S_{\rm imp}$ (solid square) and specific heat $C_{\rm imp}$ (solid triangle)
for $\lambda=0.2$, $U=1$, $W=-10^{-3}$, $x=-1.0$, and $V=0.6$.
(b) Multipole susceptibilities for the same parameters in (a).
(c) The peak temperature $T_{p2}$ (solid circle) and
the Kondo temperature $T_{\rm K}$ (solid square)
as functions of $V$.
The solid curve denotes $\Delta E_3$, whereas the horizontal broken line
indicates the local CEF excitation energy.
The dotted curve denotes the curve of $\ln T_{\rm K}=b-a/V^2$
with appropriate constants $a$ and $b$.
}
\end{figure}

Thus far, we have considered the situation with small $V$ such as $V=0.05$,
in order to focus on the localized properties of $f^7$-electron systems.
Next we consider the case with lager $V$ for the purpose to
visualize the Kondo phenomenon.
When we increase the values of $V$ for Figs.~7(a)-7(c),
we notice that the Kondo effect occurs for the case of
$\Gamma_7^-$ ground state, while for the cases of
$\Gamma_6^-$ and $\Gamma_8^-$ ground states, we cannot
observe the Kondo behavior even if we increase the magnitude of
$V$ up to $3.0$.
We remark that the $a_{\rm u}$ conduction band is hybridized only with
$\Gamma_7^-$ state,
since it includes $(f^{\dag}_{m=2,\sigma}-f^{\dag}_{m=-2,\sigma})|0\rangle /\sqrt{2}$,
while $\Gamma_6^-$ has no component of $f^{\dag}_{m=2,\sigma}|0\rangle$ and
$f^{\dag}_{m=-2,\sigma})|0\rangle$ and $\Gamma_8^-$ state contains
the component of $(f^{\dag}_{m=2,\sigma}+f^{\dag}_{m=-2,\sigma})|0\rangle /\sqrt{2}$.
\cite{Onodera}
Thus, in the following, we consider only the case of the $\Gamma_7^-$ ground state.

In Fig.~9(a), we depict the temperature dependence of entropy and specific heat
for $\lambda=0.2$, $U=1$, $W=-10^{-3}$, $x=-1.0$, and $V=0.6$.
We find the plateau of $\ln 8$ around at $T=10^{-2}$
and the entropy is changed to $\ln 2$ around at $T=10^{-3}$,
forming the peak at $T_{\rm p2}$ in the specific heat.
Then, the entropy $\ln 2$ is eventually released around
at $T=10^{-7}$, leading to the Kondo temperature $T_{\rm K}$.
The behavior of the entropy and specific heat are essentially the same as
that in Fig.~6(a) without the CEF potential.
This is nor surprising, since the level splitting due to the hybridization plays
the same role as that due to the CEF potentials.

In Fig.~9(b), we plot the multipole susceptibility
for the same parameters in Fig.~9(a).
For $T > T_{\rm K}$, we find
qualitatively the same behaviors as those in Fig.~8(c).
Around at the Kondo temperature, the Curie constant for
4u dipole gradually becomes zero, suggesting the screening
of 4u dipole moment by the conduction electrons.
Namely, the standard Kondo effect is considered to occur
in this case.

When we change the values of $V$, we plot the peak temperature
$T_{\rm p2}$ and the Kondo temperature $T_{\rm K}$ in Fig.~9(c).
Again it is found that $T_{\rm p2}$ is well scaled by $\Delta E_3$,
but we note that $\Delta E_3$ converges to the local
CEF excitation energy.
Note here that the temperature is discrete in the NRG calculation,
since it is defined as $T=\Lambda^{-(N-1)/2}$,
where $\Lambda$ is the cut-off and $N$ is the NRG step.
Thus, the peak temperature $T_{\rm p2}$ includes an error-bar
in the order of $\sqrt{\Lambda}$.
Due to the above reason, for small $V$, the solid circles seem to
scatter around the horizontal broken line.

However, for $V>1.0$, $T_{\rm p2}$ is clearly larger than the
CEF excitation energy and it seems to be scaled by $\Delta E_3$.
For small $V$, the level splitting due to the hybridization is
small and $\Delta E_3$ is almost determined by the CEF excitation 
energy.
When $V$ is increased, the hybridization effect on the level splitting
overcome the CEF potentials and $\Delta E_3$ begins to deviate from
the local CEF excitation energy.
We emphasize that $T_{\rm p2}$ correctly follow the above behavior
of $\Delta E_3$.

Finally, we discuss the Kondo temperature.
As shown by the dotted curve in Fig.~9(c),
$T_{\rm K}$ is well fitted by $b{\rm exp}(-a/V^2)$ with appropriate constants
$a$ and $b$.
This fact suggests that $T_{\rm K}$ actually indicates the Kondo temperature.
Note that for $V>1.0$, $T_{\rm K}$ becomes so high that we cannot observe
the Kondo effect in the present temperature range.

\section{Discussion and Summary}

In this paper, we have discussed the Kondo effect in $f^7$-electron systems
on the basis of the seven-orbital Anderson model by using the NRG technique.
Note that we have considered the single $a_{\rm u}$ conduction band.
We have clarified that our understanding on the Kondo effect
in the Ce compound also works in the  $f^7$-electron system
with the spin-orbit coupling.
If we are simply based on the $j$-$j$ couping scheme,
the result seems to be almost obvious.
Namely, in the limit of large spin-orbit coupling, we accommodate one
electron in $j$=7/2 octet, while $j$=5/2 sextet is fully occupied.
When one electron in the octet is hybridized with conduction band,
it is possible to understand the Kondo effect in a similar way
as that for the Ce compounds.

From the qualitative viewpoint, the $j$-$j$ coupling scenario may work
for the understanding of the Kondo effect in Eu compounds.
However, it is unclear whether the scenario is really valid for actual compounds
with the finite value of the spin-orbit coupling.
In particular, in actual rare-earth materials, the effect of Coulomb
interaction is generally stronger that that of the spin-orbit coupling.
Thus, naively speaking, it is difficult to accept even qualitatively the scenario
based on the $j$-$j$ coupling scheme in the rare-earth compounds.
It has been the motivation of the present paper to clarify this point.
Our results have clearly suggested that the Ce-compound-like Kondo phenomena
should occur for $f^7$-electron systems
in the region of $\lambda/U$ in the order of $0.1$,
which is the value of actual materials.

Since the purpose of this paper has been to clarify the appearance of
the Kondo phenomena in $f^7$-electron systems from the conceptual
viewpoint, we have not discussed the quantitative explanation of
actual Eu compounds.
In order to promote further the study of the Kondo effect in Eu compounds,
it is inevitable to include the effect of valence fluctuations,
which has been ignored in this paper.
It has been pointed out that the valence fluctuations are involved
in the heavy-electron formation in Eu compounds.\cite{Mitsuda2}
Interesting properties induced by critical valence fluctuations
have been actively investigated by Watanabe and Miyake.\cite{Miyake}

A way to investigate the effect of valence fluctuations on the Kondo phenomena
in $f^7$-electron systems is to include further the inter-site Coulomb
interaction between impurity and conduction sites.
Although for the standard impurity Anderson model without orbital degeneracy,
it is easy to include such an inter-site repulsion,
in the present seven-orbital Anderson model,
there are several possibilities to consider such orbital-dependent
inter-site repulsion.
Thus, the first step to promote the research of the effect of valence fluctuations
on the Kondo phenomena is to construct the valid model including
inter-site repulsions.
It is one of future problems.

We have considered one $a_{\rm u}$ conduction band,
but it can be validated, for instance, in cage-structure compounds
such as filled skutterudites,
since the band-structure calculations have revealed that
the main conduction band composed of pnictogen $p$ electrons
possess $a_{\rm u}$ symmetry.\cite{Harima}
If Eu ion becomes divalent in the cage of filled skutterudites,
it is interesting to investigate Eu-based filled slitterudites.
Another cage-structure can be also the candidate of the research
of the Kondo effect in $f^7$-electron systems.
If it is possible to synthesize Eu-based 1-2-20 compounds,
it may be interesting.
As for the theoretical task, it is necessary to increase the
number of conduction bands.
For instance, we consider $e_{\rm u}$
conduction bands with double degeneracy.
Since they are hybridized with $\Gamma_8^-$ states,
it may be possible to obtain the two-channel Kondo
effect in the Eu compounds.
It is another interesting future problem.

In summary, 
we have analyzed the seven-orbital Anderson model by employing
the NRG technique.
When we have included the spin-orbit coupling in the situation,
we confirm the Kondo behavior similar to that in the Ce compound
with increasing the magnitude of the spin-orbit coupling,
as expected from the $j$-$j$ coupling scheme.
An important point is that the Kondo effect
similar to the case of $n=1$ occurs for realistic values
of the spin-orbit coupling even in the case of $n=7$.
Even when we include the cubic CEF potentials,
we have also found the Kondo behavior similar to Ce compounds
for some CEF parameter regions.

\section*{Acknowledgement}

The author thanks K. Hattori, K. Kubo, Y. \=Onuki, and K. Ueda
for discussions on heavy-electron systems.
The computation in this work has been partly done using the facilities of the
Supercomputer Center of Institute for Solid State Physics, University of Tokyo.


\end{document}